\documentclass[proceedings]{JHEP37}
\usepackage{eurosym}
\usepackage{amssymb}
\usepackage{amsfonts}
\usepackage{amsmath}
\usepackage{epsfig}

\newbox\mybox

\newcommand\fverb{\setbox\mybox=\hbox\bgroup\verb}
\newcommand\fverbdo{\egroup\medskip\noindent\fbox{\unhbox\mybox}\ }
\newcommand\fverbit{\egroup\item[\fbox{\unhbox\mybox}]}
\conference{Time-dependent quasi-Hermiticity}
\abstract{It has been argued that it is incompatible to maintain unitary time-evolution for time-dependent non-Hermitian Hamiltonians when the metric operator is explicitly 
time-dependent. We demonstrate here that the time-dependent Dyson equation and the time-dependent 
quasi-Hermiticity relation
can be solved consistently in such a scenario for a time-dependent Dyson map and time-dependent metric operator, respectively. These solutions are obtained at the cost of rendering the non-Hermitian Hamiltonian to be a non-observable operator as it ceases to be quasi-Hermitian when the metric becomes
time-dependent.}

\title{Unitary quantum evolution for time-dependent quasi-Hermitian systems
with non-observable Hamiltonians}
\author{Andreas Fring$^\bullet$ and Miled H.Y. Moussa$^{\circ,\bullet}$ \\
$\bullet$ Department of Mathematics, City University London,\\
$\,\,$ Northampton Square, London EC1V 0HB, UK\\
$\circ$ Instituto de F\'{\i}sica de S\~{a}o Carlos, Universidade de S\~{a}o
Paulo, \\
$\,\,$ Caixa Postal 369, 13560-970, S\~{a}o Carlos, S\~{a}o Paulo, Brazil\\
E-mail: a.fring@city.ac.uk, miled@ifsc.usp.br}

\input{tcilatex}
\begin{document}

\section{Introduction}

The time-evolution of Hamiltonian systems is a central and fundamental issue
in quantum mechanics, especially with regard to physical applications. The
key principles are very well understood for a long time for Hermitian
Hamiltonian systems and can be found in almost any standard book on quantum
mechanics. However, the situation is quite different for the class of
non-Hermitian systems that possess real or at least partially real
eigenvalue spectra. Such type of models have been investigated sporadically
for a long time, but the relatively recent seminal paper \cite{Bender:1998ke}
has initiated a more systematic study. For time-independent systems the
governing principles are by now also well understood and many experiments
exist to confirm the key findings, e.g. \cite{Muss,MatMakris,Guo}. For
recent reviews on the subject area see for instance \cite{Benderrev,Alirev}
or \cite{BFGJ,Znojilspecial} for recent special issues.

In contrast, time-dependent non-Hermitian systems are far less well
investigated and it appears that so far no consensus has been reached about
a number of central issues. Whereas the treatment for systems with
time-dependent non-Hermitian Hamiltonians with time-independent metric
operators \cite{CA,CArev} is widely accepted the more general setting with a
time-dependent metric is still controversially discussed \cite%
{time1,time2,time3,time4,time5,time6,time7,mmaamache}. Explicit solutions to
the central equations, i.e. the time-dependent Dyson and the time-dependent
quasi-Hermiticity relation, have not been reported. Instead most authors
resort to a non-unitary time evolution \cite%
{time2,time4,time6,time7,mmaamache} for these systems by insisting on a
quasi-Hermiticity relation between a Hermitian and a non-Hermitian
\textquotedblleft Hamiltonian\textquotedblright . The main purpose of this
manuscript is to demonstrate that this is in fact not necessary. We add some
clarifying arguments to the central discussion, provide some analytic
solutions to the key equations and discuss some of the consequences.

Our manuscript is organized as follows: In section 2 we state the general
framework for a description of a unitary time-evolution for time-dependent
non-Hermitian Hamiltonians. In section 3 we provide two explicit examples
that illustrate the working of our proposal and in section 4 we state our
conclusions.

\section{The time-dependent Dyson and quasi-Hermiticity relation}

As our starting point we take the two time-dependent Schr\"{o}dinger
equations (TDSE)%
\begin{equation}
h(t)\phi (t)=i\hbar \partial _{t}\phi (t),\qquad \text{and\qquad }H(t)\Psi
(t)=i\hbar \partial _{t}\Psi (t).  \label{TS}
\end{equation}%
Both Hamiltonians involved are explicitly time-dependent, with $h(t)$ being
Hermitian whereas $H(t)$ is taken to be non-Hermitian, i.e. $h(t)=h^{\dagger
}(t)$ and $H(t)\neq H^{\dagger }(t)$. We also insist here that operators may
only be referred to as Hamiltonians if they generate the time-evolution for
the system under consideration, that is if they satisfy the TDSE. Next we
assume that the two solutions $\phi (t)$ and $\Psi (t)$ to (\ref{TS}) are
related by a time-dependent invertible operator $\eta (t)$ as%
\begin{equation}
\phi (t)=\eta (t)\Psi (t).  \label{sol}
\end{equation}%
It then follows immediately by direct substitution of (\ref{sol}) into (\ref%
{TS}) that the two Hamiltonians are allied to each other as%
\begin{equation}
h(t)=\eta (t)H(t)\eta ^{-1}(t)+i\hbar \partial _{t}\eta (t)\eta ^{-1}(t).
\label{hH}
\end{equation}%
Thus $h(t)$ and $H(t)$ are no longer related by a similarity transformation,
or more formally by the adjoint action of the Dyson operator, as in the
completely time-independent scenario \cite{dyson1956} or the time-dependent
scenario with time-independent metric, but instead their mutual dependence
involves a gauge-like term as discussed in \cite{CA,CArev,time1}. We
emphasize, however, that although formally the last term in (\ref{hH})
resembles a gauge connection this is not the role it plays here. We refer to
equation (\ref{hH}) in as the \emph{time-dependent Dyson relation} as it
generalizes its time-independent counterpart. Taking the Hermitian conjugate
of equation (\ref{hH}) and using the Hermiticity of $h(t)$ yields a relation
between $H(t)$ and its Hermitian conjugate%
\begin{equation}
H^{\dagger }(t)\eta ^{\dagger }(t)\eta (t)-\eta ^{\dagger }(t)\eta
(t)H(t)=i\hbar \partial _{t}\left[ \eta ^{\dagger }(t)\eta (t)\right] .
\label{etaH}
\end{equation}%
Interpreting $\rho (t):=\eta ^{\dagger }(t)\eta (t)$ as a metric operator
this relation replaces the standard quasi-Hermiticity relation well known in
the context time-independent non-Hermitian quantum mechanics \cite{Urubu}.
The justification for this interpretation emerges as a consistency
requirement from demanding the existence of a metric operator $\rho (t)$,
such that time-dependent probability densities in the Hermitian and
non-Hermitian system are related as 
\begin{equation}
\left\langle \phi (t)\left\vert \tilde{\phi}(t)\right\rangle \right.
=\left\langle \Psi (t)\left\vert \rho (t)\tilde{\Psi}(t)\right\rangle
\right. =:\left\langle \Psi (t)\left\vert \tilde{\Psi}(t)\right\rangle
\right. _{\rho }.  \label{prob}
\end{equation}%
For unitary time-evolution these probabilities are preserved in time such
that the derivative of both sides with respect to time must vanish. For the
left hand side this is simply guaranteed by the Hermiticity of $h(t)$ and
the validity of the corresponding TDSE (\ref{TS}). The right hand side
yields instead the consistency relation%
\begin{equation}
H^{\dagger }(t)\rho (t)-\rho (t)H(t)=i\hbar \partial _{t}\rho (t),
\label{quasi}
\end{equation}%
which when compared to (\ref{etaH}) allows for the aforementioned
identification for $\rho (t)$ in terms of $\eta (t)$ as announced above. We
refer to equation (\ref{quasi}), which may already be found in \cite{time1},
as the\emph{\ time-dependent quasi-Hermiticity relation}. It is noteworthy
to point out that the reverse statement also holds, i.e. metric operators
that do not satisfy (\ref{quasi}) do not allow for unitary time-evolution.

It is now evident that in complete analogy to the time-independent scenario
any self-adjoint operator $\emph{o(t)}$, i.e. an observable, in the
Hermitian system has an observable counterpart $\mathcal{O}(t)$~in the
non-Hermitain system related to each other as $\mathcal{O}(t)\emph{=}$ $\eta
^{-1}(t)\emph{o(t)}\eta (t)$, since%
\begin{equation}
\left\langle \phi (t)\left\vert \emph{o(t)}\tilde{\phi}(t)\right\rangle
\right. =\left\langle \emph{o(t)}\phi (t)\left\vert \tilde{\phi}%
(t)\right\rangle \right. =\left\langle \Psi (t)\left\vert \mathcal{O}(t)%
\tilde{\Psi}(t)\right\rangle \right. _{\rho }=\left\langle \mathcal{O}%
(t)\Psi (t)\left\vert \tilde{\Psi}(t)\right\rangle \right. _{\rho }.
\label{metric}
\end{equation}%
Obviously due to equation (\ref{hH}), the non-Hermitian Hamiltonian $H(t)$
does not belong to the set of observables in this system as it is not
related to $h(t)$ by a similarity transformation, which was already pointed
out in \cite{CA,CArev,time3,time5}. However, there is no compelling reason
why the non-Hermitian Hamiltonian $H(t)$ ought to be observable.
Nonetheless, one may easily find a closely related operator 
\begin{equation}
\tilde{H}(t)=\eta ^{-1}(t)h(t)\eta (t)=H(t)+i\hbar \eta ^{-1}(t)\partial
_{t}\eta (t),  \label{funkyH}
\end{equation}%
which is observable as it is related to the Hermitian observable $h(t)$ by
means of the aforementioned similarity transformation. In other words $%
\tilde{H}(t)$ is quasi-Hermitian. However, the operator $\tilde{H}(t)$ has
no obvious concrete meaning and is certainly not a Hamiltonian in the sense
that it does not generate the time-evolution in this system and does not
satisfy the original TDSE.

The relations above are directly transferred to the time-evolution
operators. Recall that for the Hermitian Hamiltonian $h(t)$, satisfying (\ref%
{TS}), the unitary time-evolution to a state $\phi (t)=u(t,t^{\prime })\phi
(t^{\prime })$ from a time $t^{\prime }$ to $t$ is governed by the
time-evolution operator%
\begin{equation}
u(t,t^{\prime })=T\exp \left[ -i\int\nolimits_{t^{\prime }}^{t}dsh(s)\right]
,  \label{utt}
\end{equation}%
satisfying%
\begin{equation}
h(t)u(t,t^{\prime })=i\hbar \partial _{t}u(t,t^{\prime }),\quad
u(t,t^{\prime })u(t^{\prime },t^{\prime \prime })=u(t,t^{\prime \prime
}),\quad \text{and \quad }u(t,t)=\mathbb{I}\text{.}
\end{equation}%
As usual $T$ denotes here time-ordering. Evidently we could replace $h(t)$
by $H(t)$ or $\tilde{H}(t)$ in (\ref{utt}), with the effect that in the
former case we no longer have a unitary time evolution and in the latter we
have a contradiction since $\tilde{H}(t)$ does not satisfy the TDSE for this
system, i.e. it is not a Hamiltonian. However, given the time-evolution
operator $u(t,t^{\prime })$ for the Hermitian system it follows
straightforwardly from (\ref{metric}) that the unitary time-evolution
operator $U(t,t^{\prime })$ for the non-Hermitian system evolving $\psi
(t)=U(t,t^{\prime })\psi (t^{\prime })$ is given by%
\begin{equation}
U(t,t^{\prime })=\eta ^{-1}(t)u(t,t^{\prime })\eta (t^{\prime }).
\end{equation}

Thus we are in complete agreement with Mostafazadeh's conclusions in \cite%
{time1,time3,time5} that for time-dependent metric operators one can not
simultaneously have a unitary time-evolution and an observable arbitrary
Hamiltonian; one can only have one or the other. The treatments in \cite%
{time2,time4,time6,time7,mmaamache} give up the possibility of a unitary
time-evolution by insisting on a quasi-Hermiticity relation between $H(t)$
and $h(t)$, hence leaving the role of the non-Hermitian operator $H(t)$ in
an obscure state. Since it does not satisfy the TDSE it remains unclear by
what kind of principle it is introduced.

Thus so far the incompatibility between the unitary time-evolution and an
observable Hamiltonian is left as a negative statement \cite%
{time1,time3,time5}, apart from the above mentioned treatments for
non-Hermitian Hamiltonians of unclear origin. It appears that no attempt has
been made to solve the relations (\ref{hH}) or (\ref{quasi}). A possible
reason is that one may insist in the observability of the Hamiltonian.
However, there is no compelling reason for such a view. In the
time-independent setting it is standard procedure to commence with
non-Hermitian Hamiltonians in terms of some auxiliary variables $x$ and $p$,
which are not observable. Here we extend this principle to the Hamiltonian
itself and treat the Hamiltonian $H(t)$ as a mere auxiliary operator, which
does, however, play the role as governing the time-evolution.

\section{Solutions to the time-dependent Dyson and quasi-Hermiticity relation%
}

It is of course vital to demonstrate that the above formulae are not empty
and can indeed be solved consistently. As in the time-independent case we
have now various options to solve these equations depending on the quantity
or quantities given at the starting point. In general, we commence with the
non-Hermitian Hamiltonian $H(t)$ satisfying the TDSE (\ref{TS}). One may
then compute, at least in principle, the metric $\rho (t)$ from the
time-dependent quasi-Hermiticity relation (\ref{quasi}) as $\rho (t)$ is the
only unknown quantity therein. The Dyson map $\eta (t)$ then follows
directly from its relation to $\rho (t)$, in which for simplicity one may
assume $\eta (t)$ to be Hermitian such that one just has to take the square
root. When $\eta (t)$ and $H(t)$ are determined one can use (\ref{hH}) to
compute directly the Hermitian counterpart $h(t)$. The final step then
consists of solving either of the TDSE (\ref{TS}) for $\phi (t)$ or $\Psi
(t) $, obtaining the counterpart simply from (\ref{sol}). Alternatively one
may also make a suitable Ansatz for $\eta (t)$ and compute the right hand
side of (\ref{hH}) demanding the result to be Hermitian. Let us see this in
detail for two examples by solving (\ref{hH}) in the first and (\ref{quasi})
in the second.

\subsection{Non-Hermitian harmonic oscillator with linear terms}

We consider first the time-dependent Hamiltonian for the harmonic oscillator
with additional linear terms in the standard creation and annihilation
operators $a$ and $a^{\dagger }$, respectively, 
\begin{equation}
H(t)=\omega (t)a^{\dagger }a+\alpha (t)a+\beta (t)a^{\dagger },\quad \omega
(t),\alpha (t),\beta (t)\in \mathbb{C}.  \label{Ham}
\end{equation}%
For convenience we set here and in what follows $\hbar =1$. Evidently $H(t)$
is non-Hermitian when $\alpha (t)\neq \beta ^{\ast }(t)$. Notice that when
demanding $\mathcal{PT}$-symmetry for the Hamiltonian in the
time-independent setting one demands $\omega (t)$, $\alpha (t)$, $\beta
(t)\rightarrow \omega ,i\alpha ,i\beta \in \mathbb{R}$, since $\mathcal{PT}:$
$a\rightarrow -a$, $a^{\dagger }\rightarrow -a^{\dagger }$. However, any
real-valued function $\omega (x)$, $i\alpha (x)$, $i\beta (x)$ may now be
replaced for instance by the complex-valued functions $\omega (it)$, $%
i\alpha (it)$, $i\beta (it)$ still leaving the Hamiltonian $\mathcal{PT}$%
-symmetric, since $\mathcal{PT}:$ $t\rightarrow -t$, $i\rightarrow -i$. In
order to solve the time-dependent Dyson relation (\ref{hH}) we make a
natural Ansatz for the time-dependent Dyson map 
\begin{equation}
\eta (t)=e^{\gamma (t)a+\lambda (t)a^{\dagger }}\quad \gamma (t),\lambda
(t)\in \mathbb{C}.  \label{Dyson}
\end{equation}%
as being similar in form to the Hamiltonian in the argument of the
exponential. Substituting $\eta (t)$ into (\ref{hH}) yields 
\begin{equation}
h(t)=\omega (t)a^{\dagger }a+u(t)a+v(t)a^{\dagger }+f(t),  \label{hermcon}
\end{equation}%
with the constraints 
\begin{equation}
u=\alpha +\omega \gamma +i\dot{\gamma},\quad \quad v=\beta -\omega \lambda +i%
\dot{\lambda},\quad \quad f=\frac{i}{2}\left( \gamma \dot{\lambda}-\dot{%
\gamma}\lambda \right) -\omega \gamma \lambda -\alpha \lambda +\beta \gamma .
\end{equation}%
As common we denote time-derivatives by an overhead dot. For $h(t)$ in (\ref%
{hermcon}) to be Hermitian we require the additional constraints $\omega
(t)\in \mathbb{R}$, $u=v^{\ast }$ and $f=f^{\ast }$, which correspond to the
two equations 
\begin{eqnarray}
\alpha -\beta ^{\ast }+\omega (\gamma +\lambda ^{\ast })+i\left( \dot{\gamma}%
+\dot{\lambda}^{\ast }\right)  &=&0,  \label{const1} \\
\frac{i}{2}\left( \gamma \dot{\lambda}-\dot{\gamma}\lambda +\gamma ^{\ast }%
\dot{\lambda}^{\ast }-\dot{\gamma}^{\ast }\lambda ^{\ast }\right) +\omega
\left( \gamma ^{\ast }\lambda ^{\ast }-\gamma \lambda \right) +\alpha ^{\ast
}\lambda ^{\ast }-\alpha \lambda +\beta \gamma -\beta ^{\ast }\gamma ^{\ast
} &=&0.  \label{const2}
\end{eqnarray}%
Attempting to solve these equations by assuming $\eta (t)$ to be the
standard displacement operator fails, as in that case we have $\gamma
=-\lambda ^{\ast }$, which by (\ref{const1}) implies that $\alpha (t)=\beta
^{\ast }(t)$ such that our supposedly non-Hermitian Hamiltonian $H(t)$
becomes Hermitian. Alternatively we may take $\gamma =\lambda ^{\ast }$ and $%
\alpha (t)=-\beta ^{\ast }(t)$, which reduces the above to the simple
constraint 
\begin{equation}
\alpha +\omega \gamma +i\dot{\gamma}=0.  \label{constrain}
\end{equation}%
Notice that this is just saying that $u$ needs to vanish. We can in fact
solve this equation by 
\begin{equation}
\gamma (t)=e^{i\chi (t)}\left[ \gamma (0)+i\int_{0}^{t}ds\alpha (s)e^{-i\chi
(s)}\right] ,  \label{gamma}
\end{equation}%
where $\chi (t):=\int_{0}^{t}ds\omega (s)$. Thus given the model defining
functions $\alpha (t)$ and $\omega (t)$ via our starting Hamiltonian $H(t)$,
we can directly compute $\gamma (t)$. For the presented solution our
Hermitian Hamiltonian turns out to be simply the harmonic oscillator with a
time-dependent frequency and overall shift. Of course there could be more
involved solutions to  (\ref{const1}) and  (\ref{const2}). The solution $%
\phi (t)$ to the TDSE for the Hermitian Hamiltonian $h(t)$ is then easily
found as a special case of the treatment in \cite{puri1979time}, such that
we have now also obtained a solution $\Psi (t)=\eta ^{-1}(t)\phi (t)$ to the
TDSE for the non-Hermitian Hamiltonian $H(t)$ subject to the above mentioned
constraints. For the convenience of the reader we recall the solution from 
\cite{puri1979time}. The ground state $\left\vert \phi _{0}(t)\right\rangle $
was found to be a coherent state $\left\vert \theta (t)\right\rangle $
dressed with a time-dependent Lewis-Riesenfeld phase $\Phi _{0}(t)$ 
\begin{equation}
\left\vert \phi _{0}(t)\right\rangle =e^{i\varphi _{0}(t)}\left\vert \theta
(t)\right\rangle ,
\end{equation}%
given by%
\begin{equation}
\left\vert \theta (t)\right\rangle =e^{-\left\vert \vartheta (t)\right\vert
^{2}}\sum\nolimits_{n=0}^{\infty }\frac{\vartheta ^{n}(t)}{\sqrt{n!}}%
\left\vert n\right\rangle ,~~~\vartheta (t)=\vartheta (0)e^{-i\chi
(t)},~~\varphi _{0}(t)=\varphi _{0}(0)-\int_{0}^{t}dsf(s),
\end{equation}%
with $\left\vert n\right\rangle $ being a standard Fock eigenstate of the
number operator $a^{\dagger }a$. Excited states are constructed in a similar
fashion, see also \cite{Miled1,Miled2} for further details.

The observables in the non-Hermitian system are easily computed. For
instance, the quadratures $(X,P)$ corresponding in the Hermitian system to
the coordinate and momentum operators $x=\left( a^{\dagger }+a\right) /\sqrt{%
2}$ and $p=i\left( a^{\dagger }-a\right) /\sqrt{2}$, respectively, are now
simply shifted operators in the original variables%
\begin{equation}
X=\eta ^{-1}x\eta =x-i\sqrt{2}\func{Im}\gamma ,\quad \text{and\quad }P=\eta
^{-1}p\eta =p-i\sqrt{2}\func{Re}\gamma .
\end{equation}%
The observable operator related to the Hermitian Hamiltonian, albeit not
satisfyimg the original TDSE, results to%
\begin{equation}
\tilde{H}(t)=\eta ^{-1}(t)h(t)\eta (t)=\omega (t)\left[ a^{\dagger }a-\gamma
(t)a+\gamma ^{\ast }(t)a^{\dagger }\right] +\frac{i}{2}\left[ \dot{\gamma}%
(t)\gamma ^{\ast }(t)-\gamma (t)\dot{\gamma}^{\ast }(t)\right] .
\end{equation}%
We notice that $\tilde{H}(t)$ and $H(t)$ have the same structure in their
operator content.

\subsection{Non-Hermitian spin chain}

Next we consider a discretised lattice version of the Yang-Lee model
proposed originally in \cite{gehlen1}. The model is an Ising quantum spin
chain in the presence of a magnetic field in the $z$-direction together with
a longitudinal imaginary field in the $x$-direction 
\begin{equation}
H_{N}(t)=-\frac{1}{2}\sum_{j=1}^{N}(\sigma _{j}^{z}+\lambda (t)\sigma
_{j}^{x}\sigma _{j+1}^{x}+i\kappa (t)\sigma _{j}^{x}),\quad \lambda
(t),\kappa (t)\in \mathbb{C}.  \label{H}
\end{equation}%
The boundary conditions for the Pauli spin matrices are taken to be $\sigma
_{1}=\sigma _{N+1}$. Here we modify the model by introducing a
time-dependence into the coupling constants by replacing $\lambda $, $\kappa 
$ in previous studies by time-dependent functions $\lambda (t)$, $\kappa (t)$%
. The $\mathcal{PT}$-symmetry of the Hamiltonian is $\mathcal{PT}:$ $\sigma
^{x}\rightarrow -\sigma ^{x}$, $\sigma ^{z}\rightarrow \sigma ^{z}$, $%
t\rightarrow -t$, $i\rightarrow -i$. For small length $N$ time-independent
Dyson maps, metric operators and isospectral counterparts have been
constructed in \cite{chainOla}. We present here the simplest example for the
time-dependent scenario by taking $N=1$, such that the Hamiltonian acquires
the form of a simple non-Hermitian $2\times 2$-matrix%
\begin{equation}
H_{1}(t)=-\frac{1}{2}\left[ \sigma _{1}^{z}+\lambda (t)\sigma _{1}^{x}\sigma
_{1}^{x}+i\kappa (t)\sigma _{1}^{x}\right] =-\frac{1}{2}\left( 
\begin{array}{cc}
1+\lambda (t) & i\kappa (t) \\ 
i\kappa (t) & \lambda (t)-1%
\end{array}%
\right) .
\end{equation}%
Instead of solving equation (\ref{hH}) as in the previous subsection, we now
attempt here to solve the time-dependent quasi-Hermiticity relation (\ref%
{quasi}) for the metric operator $\rho (t)$ by assuming the most general
Hermitian form as an Ansatz 
\begin{equation}
\rho (t)=\left( 
\begin{array}{cc}
\alpha (t) & \beta (t)+i\gamma (t) \\ 
\beta (t)-i\gamma (t) & \delta (t)%
\end{array}%
\right) ,\qquad \alpha (t),\beta (t),\gamma (t),\delta (t)\in \mathbb{R}.
\end{equation}%
Taking $\lambda (t),\kappa (t)\in \mathbb{R}$, the substitution of $\rho (t)$
into (\ref{quasi}) yields

\begin{equation}
\left( 
\begin{array}{cc}
\dot{\alpha}-\beta \kappa & \gamma -\frac{\kappa }{2}(\alpha +\delta )+\dot{%
\beta}+i\dot{\gamma}-i\beta \\ 
\gamma -\frac{\kappa }{2}(\alpha +\delta )+\dot{\beta}+i\beta -i\dot{\gamma}
& \dot{\delta}-\beta \kappa%
\end{array}%
\right) =0.
\end{equation}%
The equations resulting from each matrix entry are solved by%
\begin{equation}
\alpha (t)=\alpha _{0}+\int\nolimits_{0}^{t}ds\beta (s)\kappa (s),\quad
\delta (t)=\delta _{0}+\int\nolimits_{0}^{t}ds\beta (s)\kappa (s),\quad
\gamma (t)=\gamma _{0}+\int\nolimits_{0}^{t}ds\beta (s),  \label{entry}
\end{equation}%
with $\beta (t)$ constraint to 
\begin{equation}
\dot{\beta}(t)+\int\nolimits_{0}^{t}ds\beta (s)-\kappa
(t)\int\nolimits_{0}^{t}ds\beta (s)\kappa (s)-\frac{\kappa (t)}{2}(\alpha
_{0}+\delta _{0})+\gamma _{0}=0.  \label{constraint}
\end{equation}%
The latter equation is nontrivial, but we will demonstrate that it actually
possesses meaningful solutions. A great simplification is achieved by
assuming $\beta (t)=\dot{\kappa}(t)$, since then the two integrals may be
solved easily, leaving us with a second order differential equation for the
time-dependent function $\kappa (t)$%
\begin{equation}
\ddot{\kappa}(t)+\kappa (t)\left( 1-\frac{\alpha _{0}+\delta _{0}}{2}+\frac{%
\kappa ^{2}(0)}{2}\right) -\frac{1}{2}\kappa ^{3}(t)+\gamma _{0}-\kappa
(0)=0.  \label{kappa}
\end{equation}%
Given the values for the entries in the matrix $\rho $ as in (\ref{entry}),
with the above assumption and implementing (\ref{kappa}) we find an
additional constraint on the combination of initial values%
\begin{equation}
\left\vert \rho (t)\right\vert =\frac{1}{4}\left[ \kappa ^{2}(0)-2\alpha _{0}%
\right] \left[ 2\delta _{0}-\kappa ^{2}(0)\right] -\left[ \gamma _{0}-\kappa
^{2}(0)\right] >0,
\end{equation}%
to guarantee a positive definite metric.

In general solution to (\ref{kappa}) are Jacobi elliptic functions, that is
complex, which are however excluded by the fact that $\alpha (t)\,$, $\beta
(t)$, $\gamma (t)\,\ $and $\delta (t)$ have to be real by assumption.
Nonetheless, for special values of the elliptic modulus we may also obtain
several real solutions. For instance, 
\begin{eqnarray}
\kappa (t) &=&2\tan (t),\qquad \text{with ~~~}\gamma _{0}=0,\quad \alpha
_{0}=6-\delta _{0},\quad \left\vert \rho (t)\right\vert =-4+6\delta
_{0}-\delta _{0}^{2},  \label{s1} \\
\kappa (t) &=&2\sec (t),\qquad \text{with ~~~}\gamma _{0}=2,\quad \alpha
_{0}=4-\delta _{0},\quad \left\vert \rho (t)\right\vert =-4+4\delta
_{0}-\delta _{0}^{2},  \label{s2} \\
\kappa (t) &=&2\tanh (t),~~~~\text{with ~~~}\gamma _{0}=0,\quad \alpha
_{0}=-2-\delta _{0},\quad \left\vert \rho (t)\right\vert =-4-2\delta
_{0}-\delta _{0}^{2},  \label{s3}
\end{eqnarray}%
solve the constraining equation (\ref{kappa}) with $\delta _{0}$ left as a
free parameter. We observe that not all of these solutions are permissible
as (\ref{s2}) and (\ref{s3}) will always lead to nonpositive operators $\rho
(t)$. However, solution (\ref{s1}) admits the possibility $\left\vert \rho
(t)\right\vert >0$ in the range $3-\sqrt{5}<\delta _{0}<$ $3+\sqrt{5}.$ For
convenience, we take now $\delta _{0}=1$ in what follows and analyze this
solution further. Using the above values, the time-dependent metric operator
is computed to 
\begin{equation}
\rho (t)=\left( 
\begin{array}{cc}
5+2\tan ^{2}(t) & 2\sec ^{2}(t)+2i\tan (t) \\ 
2\sec ^{2}(t)-2i\tan (t) & 1+2\tan ^{2}(t)%
\end{array}%
\right) ,
\end{equation}%
such that $\left\vert \rho (t)\right\vert =1$. Assuming the Dyson operator
to be Hermitian we may compute it by first diagonalizing $\rho (t)=\eta
^{2}(t)=UDU^{-1}$, with $D$ being a diagonal matrix, and subsequently
computing $\sqrt{\rho (t)}=\eta (t)=U\sqrt{D}U^{-1}$. As $\rho (t)$ is
positive definite this operation is well-defined. In this manner we obtain
the time-dependent Dyson operator 
\begin{equation}
\eta (t)=\frac{1}{\sqrt{\sec ^{2}(t)+1}}\left( 
\begin{array}{cc}
2+\sec ^{2}(t) & \sec (t)(\sec (t)+i\sin (t)) \\ 
\sec (t)(\sec (t)-i\sin (t)) & \sec ^{2}(t)%
\end{array}%
\right) .
\end{equation}%
These expressions allows us to compute the Hermitian Hamiltonian $h(t)$ by
means of (\ref{hH}) 
\begin{equation}
h(t)=\frac{1}{3+\cos (2t)}\left( 
\begin{array}{cc}
-\frac{1}{2}\left[ 1+3\lambda (t)+[3+\lambda (t)]\cos (2t)\right] & -i\sin
(2t) \\ 
i\sin (2t) & \frac{1}{2}\left[ 1-3\lambda (t)+[3-\lambda (t)]\cos (2t)\right]%
\end{array}%
\right) .
\end{equation}%
Evidently there might be many more solutions when allowing $\lambda
(t),\kappa (t)$ to have nonvanishing imaginary parts or when relaxing the
assumption on $\beta (t)$ in solving (\ref{constraint}). Here it suffices to
demonstrate that some meaningful solutions exists.

\section{Conclusions}

We have demonstrated that the time-dependent quasi-Hermiticity relations (%
\ref{quasi}) and therefore also the time-dependent Dyson relation (\ref{hH})
possess meaningful solutions. This means a consistent description of a
unitary quantum time-evolution with time-dependent metric is indeed
possible. Unlike as in previous treatments we do not demand a
quasi-Hermiticity relation between a Hermitian Hamiltonian and a
non-Hermitian Hamiltonian, which inevitably leads to non-unitary quantum
evolution. Instead, we do not demand the observability of the non-Hermitian
Hamiltonian that satisfies the TDSE and simply treat it as an auxiliary
operator. Nonetheless, the system still possess a well-defined observable
Hamiltonian in form of $h(t)$.

Evidently there are still many open problems. Clearly more explicit
solutions for concrete models would shed further light on the viewpoint we
proposed. The uniqueness problem of the metric operator in the
time-independent case is well known, i.e. given a non-Hermitian Hamiltionian
as a starting point of the construction one obtains numerous consistent
solutions for the metric operator. This issue is still unresolved to a large
extent in the time-independent scenario. For the time-dependent case this
difficulty appears to be much more amplified and solutions are even more
ambiguous. However, more complex settings often allow to find special
criteria for very particular solutions and the hope is that one might be
able to extract concrete selection criteria from these considerations.

\bigskip \noindent \textbf{Acknowledgments:} MHYM would like to thank CAPES,
Brazil financial agency, for support and City University London for kind
hospitality.

\newif\ifabfull\abfulltrue

\end{document}